\documentclass[aps,prb,twocolumn,a4,superscriptaddress]{revtex4}

\usepackage{amsmath,amssymb,amsfonts,upgreek}
\usepackage{color}
\usepackage{graphicx}

\begin{document}

\title{Rubidium superoxide: a $p$-electron Mott insulator}

\date{\today}

\author{Roman Kov\'a\v{c}ik}
\email{r.kovacik@fz-juelich.de}
\affiliation{Peter Gr\"unberg Institut and Institute for Advanced
  Simulation, Forschungszentrum J\"ulich and JARA, 52425 J\"ulich, Germany}
\altaffiliation[Previous address: ]{School of Physics, Trinity College
  Dublin, Dublin 2, Ireland}
\author{Philipp Werner}
\affiliation{University of Fribourg, Department of Physics, Ch. du
  Mus\'ee 3, 1700 Fribourg, Switzerland}
\author{Krzysztof Dymkowski}
\affiliation{Materials Theory, ETH Z\"urich, Wolfgang-Pauli-Strasse
  27, 8093 Z\"urich, Switzerland}
\altaffiliation[Previous address: ]{School of Physics, Trinity College
  Dublin, Dublin 2, Ireland}
\author{Claude Ederer}
\email{claude.ederer@mat.ethz.ch}
\affiliation{Materials Theory, ETH Z\"urich, Wolfgang-Pauli-Strasse
  27, 8093 Z\"urich, Switzerland}
\altaffiliation[Previous address: ]{School of Physics, Trinity College
  Dublin, Dublin 2, Ireland}

\begin{abstract}
Rubidium superoxide, RbO$_2$, is a rare example of a solid with
partially-filled electronic $p$ states, which allows to study the
interplay of spin and orbital order and other effects of strong
electronic correlations in a material that is quite different from the
conventional $d$ or $f$ electron systems. Here we show, using a
combination of density functional theory (DFT) and dynamical
mean-field theory, that at room temperature RbO$_2$ is indeed a
paramagnetic Mott insulator. We construct the metal-insulator phase
diagram as a function of temperature and Hubbard interaction
parameters $U$ and $J$. Due to the strong particle-hole asymmetry of
the RbO$_2$ band-structure, we find strong differences compared to a
simple semi-elliptical density of states, which is often used to study
the multiband Hubbard model. In agreement with our previous DFT study,
we also find indications for complex spin and orbital order at low
temperatures.
\end{abstract}


\maketitle

\section{Introduction}

\begin{figure}[]
  \centering
  \includegraphics[width=0.9\columnwidth]{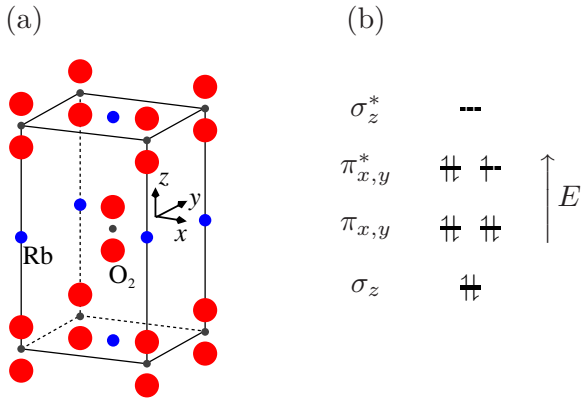}
  \caption{(a) Tetragonal crystal structure of RbO$_2$ at room
    temperature. (b) Electronic structure represented by a set of
    oxygen $p$ molecular orbitals.}
  \label{fig:as-es}
\end{figure}

Rubidium superoxide, RbO$_2$, is an interesting example of a material,
where spin and orbital order appears not as a result of partially
filled $d$ or $f$ states, but due to partially filled $p$ electron
states. RbO$_2$ is a member of the family of alkali superoxides
$A$O$_2$ ($A$ = K, Rb, or Cs), which are insulating crystalline
materials composed of $A^+$ and (O$_2$)$^-$ ions
\cite{1974_zumsteg,1979_labhart}. At room temperature, RbO$_2$ has a
tetragonal crystal structure (see Fig.~\ref{fig:as-es}(a)), while with
decreasing temperature this structure undergoes several weak
distortions, first to orthorhombic, then to monoclinic symmetry
\cite{1979_labhart,1978_rosenfeld}. The electronic structure around
the Fermi level is dominated by oxygen $p$ states which can be well
approximated by molecular orbitals (MOs) corresponding to the O$_2$
units, and are filled with 9 electrons (see Fig.~\ref{fig:as-es}(b)).
Assuming no further symmetry breaking, the two highest occupied
antibonding $\pi^*$ orbitals are 3/4 filled.

The degeneracy of these orbitals can be lifted through either magnetic
or orbital long range order, or both. The alkali superoxides thus
allow to study ``correlation effects'' in a completely different class
of materials compared to the more conventional transition metal oxides
or $f$ electron systems. Antiferromagnetic order is indeed found
experimentally at low temperatures ($T_\text{N}(\text{RbO}_2)\approx
15$~K) \cite{1974_zumsteg,1979_labhart}, and it was suggested by
recent density functional theory (DFT) and model studies that the
insulating character of alkali-superoxides at low temperatures can be
explained by the interplay of correlation effects (spin and orbital
order) and crystal distortions
\cite{2008_solovyev,2009_kovacik,2010_kim,2010_ylvisaker,2010_nandy,2011_wohlfeld}. However,
the nature of the insulating state of these superoxides at room
temperature has so far remained unexplored.

Due to the high symmetry crystal structure with no long-range order of
spins or orbitals, it is impossible to explain the insulating
character of the alkali superoxides at room temperature within an
effective single particle band picture. Here we show, using a
combination of DFT and dynamical mean field theory (DFT+DMFT), that
RbO$_2$ at room temperature is in fact a \emph{Mott insulator}, where
the strong Coulomb repulsion prevents the electron hopping between
adjacent sites.

\begin{figure}[]
  \centering \includegraphics[width=0.95\columnwidth]{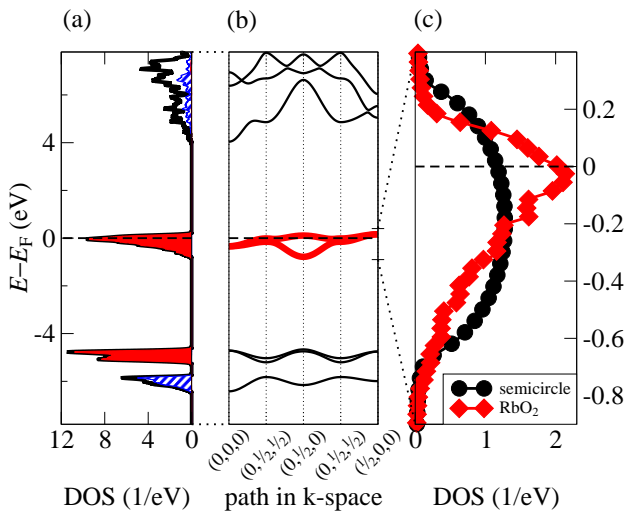}
  \caption{(Color online) Density of states (DOS) (a) and
    bandstructure (b) of nonmagnetic RbO$_2$. In (a) the total DOS is
    shown as thick (black) line, while the projection on O$_{2}$
    $\sigma$ and $\pi$ states are shown as (blue) striped and (red)
    filled areas, respectively. Panel (c) shows the DOS (per
    spin-orbital) of the $\pi^*$ bands used in the DMFT calculations,
    compared to a semi-circular DOS with the same bandwidth (a small
    broadening is applied in both cases).}
  \label{fig:dos}
\end{figure}

\section{Electronic structure of RbO$_2$}

We obtain the electronic structure of RbO$_{2}$ from a
non-spinpolarized DFT calculation using the Quantum-ESPRESSO package
\cite{quantum-espresso}, employing the generalized gradient
approximation of Perdew, Burke, and Ernzerhof \cite{1996_perdew} and
ultrasoft pseudopotentials \cite{1990_vanderbilt}.
Figs.~\ref{fig:dos}(a) and (b) show the resulting density of states
(DOS) and bandstructure. It can be seen that the electronic structure
of RbO$_2$ indeed closely resembles the simple MO picture sketched in
Fig.~\ref{fig:as-es}(b), with a splitting of about 5\,eV between the
bonding and antibonding $\pi$ and $\pi^*$ bands, and a single band
corresponding to bonding $\sigma$ MOs at $-6$\,eV. The antibonding
$\sigma^*$ states at $\sim5$\,eV are strongly intermixed with other
empty states corresponding to the Rb$^+$ cations.

\section{Dynamical mean-field theory --- Computational method}

To calculate the electronic properties at finite temperature and
account for local correlation effects, we use dynamical mean-field
theory \cite{1996_georges} (DMFT) which allows to map the lattice
problem to an effective problem of a single-site impurity surrounded
by a bath. The interaction part of the impurity Hamiltonian is taken
to be of the Slater-Kanamori form
\begin{align}
&H_\text{int}=\sum_{a} U n_{a,\uparrow} n_{a,\downarrow}+\sum_{a\ne
    b,\sigma} U' n_{a,\sigma} n_{b,-\sigma} \nonumber\\ &+\sum_{a\ne
    b,\sigma} (U'-J) n_{a,\sigma}n_{b,\sigma}\nonumber\\ &-\sum_{a\ne
    b}J(d^\dagger_{a,\downarrow}d^\dagger_{b,\uparrow}d_{b,\downarrow}d
  _{a,\uparrow} +
  d^\dagger_{b,\uparrow}d^\dagger_{b,\downarrow}d_{a,\uparrow}d_{a,\downarrow}
  + h.c.),
\label{H_int}
\end{align}
with $d^\dagger_{a,\sigma}$ the creation operator for an electron of
spin $\sigma$ in orbital $a$ and $U'=U-2J$.
To solve the effective impurity problem, we use the strong-coupling
continuous time quantum Monte Carlo approach (CT-HYB)
\cite{2006_werner, 2011_gull}. From the self-consistently determined
hybridization function $\Delta(\tau)$, the impurity Green's function
$G_\text{imp}(\tau)$ is computed and measured on a homogeneous grid of
$N_\tau=1000 \cdot \left[ \sqrt{\beta/40\,\text{eV}^{-1}}\right]$
points, where $\left[ \dots \right]$ represents the nearest integer
number. After Fourier transformation we obtain the self-energy in
Matsubara space,
\begin{equation}
\label{eq:dyson}
\Sigma(\text{i}\omega_n)=\text{i}\omega_n+\mu-G_\text{imp}^{-1}(\text{i}\omega_n)-\Delta(\text{i}\omega_n),
\end{equation}
where $\omega_n=(2n+1)\pi/\beta$ for integer $n$, $\mu$ is the
chemical potential, and $\beta=1/T$ the inverse temperature. Using
this self-energy and the single-particle Hamiltonian $H(\mathbf{k})$
we obtain the local lattice Green's function by averaging over the
Brillouin zone:
\begin{equation}\label{eq:gloc}
  G_\text{loc}(\text{i}\omega_n)=\frac{1}{N_k}\sum_{\mathbf{k}}\left[
    \text{i}\omega_n+\mu-H(\mathbf{k})-\Sigma(\text{i}\omega_n)
    \right]^{-1}\,.
\end{equation}
The DMFT self-consistency condition demands that this local lattice
Green's function is the same as the impurity Green's function.  This
condition, in combination with Eq.~(\ref{eq:dyson}) yields the
hybridization function for the next DMFT iteration,
\begin{equation}\label{eq:selfconsistency}
  \Delta(\text{i}\omega_n)=\text{i}\omega_n+\mu-G_\text{loc}^{-1}(\text{i}\omega_n)-\Sigma(\text{i}\omega_n).
\end{equation}

We only include the partially-filled antibonding $\pi^{*}_{x/y}$ bands
in our DMFT calculations for RbO$_2$, and express the corresponding
Hamiltonian $H(\mathbf{k})$ in a basis of maximally localized Wannier
functions \cite{2008_mostofi}. The corresponding DOS is shown in
Fig.~\ref{fig:dos}(c). One can recognize a pronounced asymmetry with
respect to half-filling. In DMFT studies, a model semicircle (SC)
density of states (DOS) is often employed to represent the electronic
bands, since it leads to a simple expression connecting $\Delta$ and
$G_\text{imp}$. Furthermore, due to the resulting particle-hole
symmetry, only occupations between zero and half-filling need to be
studied. Here, we investigate the differences between results obtained
using the model SC DOS and the realistic DFT band structure of
RbO$_{2}$ in the tetragonal crystal structure. The bandwidth of the SC
DOS is set equal to the bandwidth of the RbO$_{2}$ DOS (0.93~eV, see
Fig.~\ref{fig:dos}(c)).

\section{Results}

\subsection{Room temperature properties}

\begin{figure}[]
  \centering
  \includegraphics[width=0.8\columnwidth]{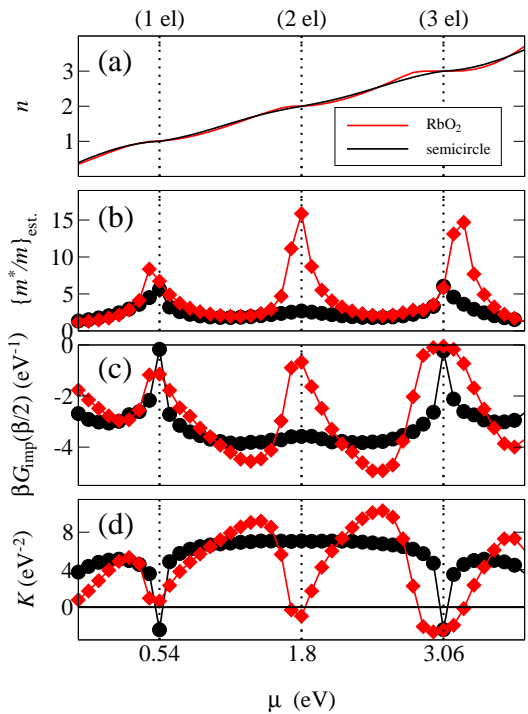}
  \caption{Various quantities evaluated from the impurity Green's
    function indicating the metal-insulator transition at integer
    filling, calculated for $\beta=40$~eV$^{-1}$ ($T\approx 290$~K),
    $U=1.2$~eV, and $J=0.0$~eV. The average over all spin-orbitals is
    shown in panels (b), (c), and (d).}
  \label{fig:dmft-12-00}
\end{figure}

Generally, the value of the spectral function at zero energy indicates
whether a material is insulating or metallic. However, obtaining the
spectral function from the imaginary time Green's function requires an
analytic continuation to the real axis, which can introduce additional
uncertainties \cite{Jarrell/Gubernatis:1996}. We therefore consider
several possible indicators for the metal-insulator transition (MIT)
which are directly accessible from $G_\text{imp}(\tau)$. All of these
quantities (described in more detail below) are compared in
Fig.~\ref{fig:dmft-12-00}, calculated at room temperature
($\beta=40$~eV$^{-1}$) for interaction parameters close to the MIT
($U=1.2$~eV and $J=0$).

One possibility is to monitor the occupation
\mbox{$n=-\sum_{\alpha}G_\text{imp}^{\alpha}(\beta)$} ($\alpha$ is the
spin-orbital index) as a function of the chemical potential $\mu$ (see
Fig.~\ref{fig:dmft-12-00}(a)) and to identify the insulating phase by
a plateau in $n(\mu)$. This, however, requires a large number of
calculations for slightly different values of $\mu$. Another indicator
is given by the mass enhancement in the low-temperature metallic
phase, which grows rapidly as the Mott insulating state is approached,
and which we estimate from the self-energy at the lowest Matsubara
frequency (see Ref.~\onlinecite{2011_fuchs}):
\begin{equation}
  \left\{\frac{m^{*}}{m}\right\}_\text{est.}=
  1-\frac{\Im[\Sigma(\text{i}\omega_0)]}{\omega_0} \quad .
\end{equation} 
We also consider the following estimate of the spectral function (see
e.g. Ref.~\onlinecite{2011_fuchs}):
\begin{equation}
  A(0) \approx -\frac{\beta}{\pi}\,
  G_\text{imp}\left(\frac{\beta}{2}\right) \quad .
\end{equation}
While for both of these quantities (shown in
Fig.~\ref{fig:dmft-12-00}(b) and~\ref{fig:dmft-12-00}(c),
respectively) only one calculation at the correct $\mu$ value is
required, the identification of the MIT phase boundary requires the
definition of a suitable threshold value. In addition, $G(\beta/2)$
suffers from significant statistical noise in the insulating state, as
this $\tau$-region is difficult to sample with standard CT-HYB. A
quantity which is quite insensitive to noise is the slope of
$G_\text{imp}(\text{i}\omega\rightarrow{0})$, which is
positive/negative for the metallic/insulating state \cite{2011_fuchs}.
In practice, we estimate the slope $K$ from
$G_\text{imp}(\text{i}\omega)$ at the two lowest Matsubara
frequencies:
\begin{equation}
  K=\left\{\frac{\text{d}\Im[{G_\text{imp}}(0)]}{\text{d}\omega}\right\}_\text{est.}
  =\frac{
    \Im[G_\text{imp}(\text{i}\omega_1)-G_\text{imp}(\text{i}\omega_0)]
  } {\omega_1-\omega_0}.
\end{equation}

In Fig.~\ref{fig:dmft-12-00}, a clear difference can be seen between
the SC and the RbO$_{2}$ input. For the SC, the insulating state
already appears at 1 and 3~el filling, while 2~el filling is clearly
metallic. In contrast, the RbO$_{2}$ electronic structure yields a
clear insulating state at 3~el filling and a (just barely) insulating
state at 2~el filling, while 1~el filling is still metallic. The
different range of $\mu$ which leads to the insulating state for SC
and RbO$_2$, respectively, indicates a sizable shift of the
corresponding MIT boundary. The particle-hole asymmetry in the real
electronic structure of RbO$_2$ thus leads to large quantitative
changes compared to the simple SC DOS.

\begin{figure}[]
 \centering
 \includegraphics[width=\columnwidth]{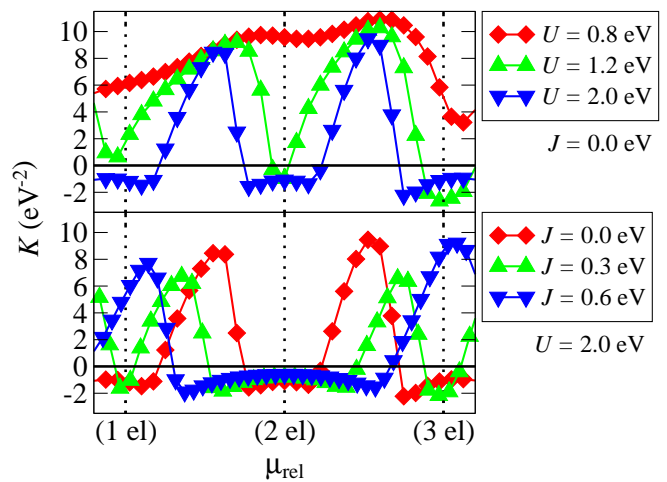}
 \caption{Slope $K$ for RbO$_{2}$ input and selected values of $U$ and
   $J$ at $\beta=40$~eV$^{-1}$ (average over all spin-orbitals is
   shown).}
 \label{fig:dmft-Mu-UJ}
\end{figure}

Fig.~\ref{fig:dmft-Mu-UJ} shows the slope $K$ for different values of
$U$ and $J$ at $T\approx{290}$~K using the RbO$_2$ band structure. For
fixed $J=0$~eV, there is an obvious tendency towards the insulating
state with increasing $U$ for all integer fillings, as
expected. However, at fixed $U=2$~eV, increasing $J$ favors the Mott
insulator at half-filling (2~el) but favors the metallic solution for
1 and 3~el fillings. This is consistent with previous discussions of
multi-orbital models \cite{2009_Werner,2011_Medici}: In the large-$U$
limit, the width of the ``Mott plateau'' in $\mu$ is given by
$\Delta^\text{Mott}_n=E_{n+1}+E_{n-1}-2E_{n}$, with $E_n$ denoting the
lowest eigenvalue of the $n$-particle eigenstates of $H_\text{int}$
(Eq.~(\ref{H_int})) \cite{2009_Werner}. In our two-orbital case this
estimate yields $U-3J$ for $n=1$, $3$ and $U+J$ for $n=2$, in
agreement with the observed dependence of the plateau-width on $J$. In
reality, the Mott plateau will be reduced by approximately the
bandwidth $W$, so that we obtain the rough estimate
\begin{equation}
\label{eq:Mott-gap}
\Delta^\text{Mott}_{3}\approx U-3J-W \quad .
\end{equation}

\begin{figure}[]
  \centering
  \includegraphics[width=0.9\columnwidth]{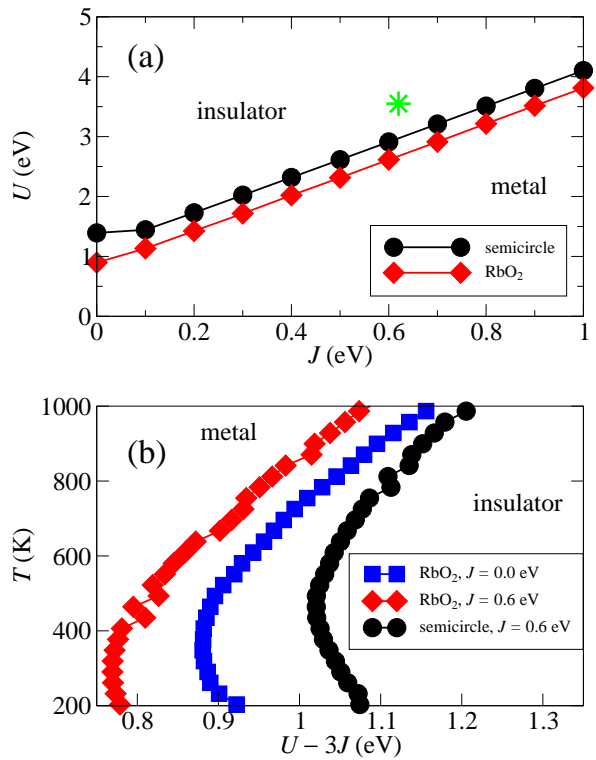}
  \caption{Metal-insulator phase diagrams. (a) Critical $U$ as
    function of $J$ at room temperature ($T\approx 290$~K). The green
    star indicates the realistic values for $U$ and $J$ calculated in
    \cite{2008_solovyev}. (b) Critical temperature as function of
    $U-3J$.}
  \label{fig:dmft-pd}
\end{figure}

Based on the identification of the MIT boundary using the slope $K$,
we computed the phase diagram for 3~el filling at room temperature
($T\approx 290$~K) as a function of the interaction parameters $U$ and
$J$ (see Fig.~\ref{fig:dmft-pd}(a)). Since $\Delta_\text{Mott}$ must
be larger than zero for a Mott insulating solution to exist,
Eq.~(\ref{eq:Mott-gap}) also provides a crude estimate for the
critical interaction strength: $U_c - 3J \approx \text{const}$. It can
be seen that the MIT boundary in Fig.~\ref{fig:dmft-pd}(a) agrees
nicely with this simple estimate. Furthermore, the critical $U-3J$ at
room temperature differs by $\approx 30\%$ of the bandwidth between
RbO$_2$ and the simple SC DOS, and increases slightly as a function of
$T$ for $T \gtrapprox 300-400$~K, while the opposite trend is observed
at lower $T$ (Fig.~\ref{fig:dmft-pd}(b)).

For realistic values of the interaction parameters $U=3.55$~eV and
$J=0.62$~eV, which were obtained for $\pi^*$ orbitals in the very
similar material KO$_{2}$ using the constrained LDA and random-phase
approximation \cite{2008_solovyev}, the insulating state is obtained
for both SC and RbO$_{2}$. Our results therefore predict a Mott
insulating state (without long-range order) for RbO$_2$ at room
temperature, consistent with experimental observations.

\begin{figure}[]
\includegraphics*[width=\columnwidth]{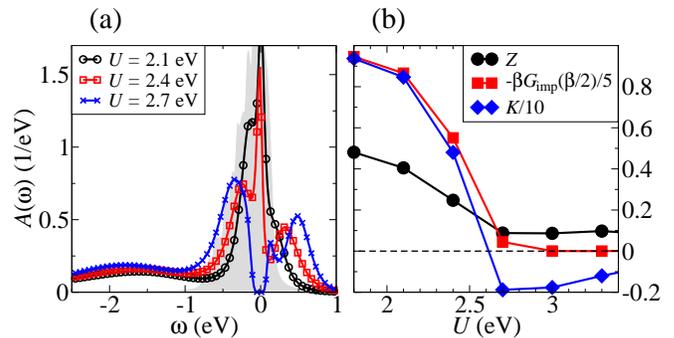}
\caption{Spectral functions (a) and the various indicators for the MIT
  (b) for RbO$_2$ at $\beta=40$\,eV$^{-1}$ ($T \approx 290$\,K),
  $J=0.6$\,eV, and different values of $U$ around the
  MIT. $Z=(m^*/m)_\text{est.}^{-1}$ and $K$ in eV$^{-2}$. The
  noninteracting DOS is shown as grey shaded area in (a).}
\label{fig:spectra}
\end{figure}

While we did not find evidence for a coexistence region, indicative of
a first order MIT, for $T \geq 145$~K, we have verified that we obtain
an insulating state with a clear gap in the spectral function, even at
room temperature. To demonstrate this, we have used the maximum
entropy method \cite{Jarrell/Gubernatis:1996} to construct spectral
functions at $\beta=40$\,eV$^{-1}$, $J=0.6$\,eV, and different values
of $U$ around the MIT. The result is shown in
Fig.~\ref{fig:spectra}(a). A gap is present for $U \geq 2.7$\,eV, in
perfect agreement with the various indicators of the MIT discussed
previously (and which are shown in Fig.~\ref{fig:spectra}(b)). In
agreement with Ref.~\onlinecite{Medici/Mravlje/Georges:2011} we find a
``bad metal'' region with a strongly renormalized $Z=(m^*/m)^{-1}
\lesssim 0.4$ in the vicinity of the Mott transition. From
Fig.~\ref{fig:spectra}(a) it can be seen that this corresponds to
spectral functions with substantial narrowing of the central
quasiparticle feature and an emerging three-peak structure visible for
$U=2.4$\,eV. In addition, there is a significant spectral weight
transfer to energies around $-2$\,eV compared to the noninteracting
DOS.

\subsection{Low temperature behavior}

\begin{figure}[]
  \centering
  \includegraphics[width=0.95\columnwidth]{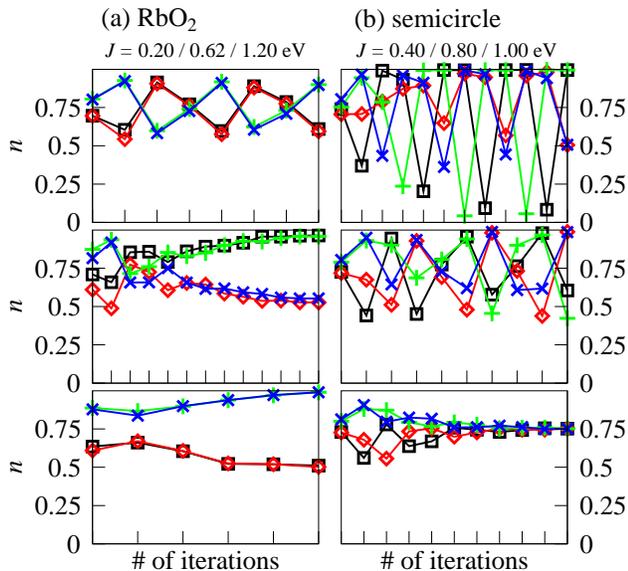}
  \caption{(Color online) Evolution of individual occupations for
    different $J$ and fixed $U-J=2.93$\,eV at $T\approx{29}$~K for
    RbO$_2$ input (a) and for the SC DOS (b). (Black) Squares: orbital
    1, spin up; (red) diamonds: orbital 2, spin up; (green) plus
    symbols: orbital 1, spin down; (blue) crosses: orbital 2, spin
    down.}
  \label{fig:dmft-occiter}
\end{figure}

Finally, we focus on the low-temperature behavior. While for
temperatures $T\geq 200$~K (for which we didn't find indications of
ordered states), the hybridization function is averaged over all
spin-orbitals in each iteration, Fig.~\ref{fig:dmft-occiter} shows the
evolution of the occupation of each individual spin-orbital at $T
\approx 29$~K when no such averaging is performed. While for some
values of $J$ the occupation eventually converges to spin and/or
orbitally polarized states, the occupations exhibits characteristic
oscillations for other values of $J$. As discussed in
\cite{2009_chan}, such oscillations indicate that the system wants to
adopt an ordered state with a sublattice structure that is
incompatible with the applied self-consistency condition (in our case
all sites are forced to be equivalent).

\begin{figure}[]
  \centering \includegraphics[width=0.8\columnwidth]{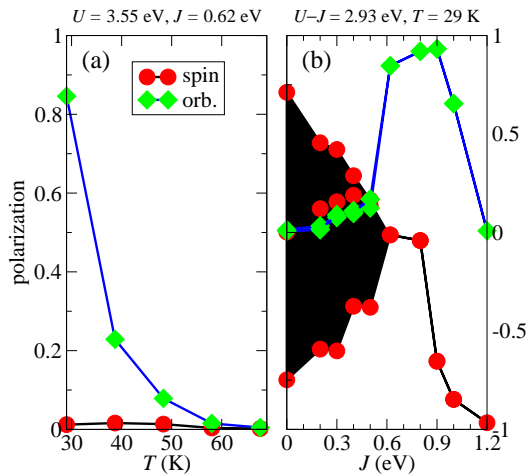}
  \caption{(a) Temperature dependence of spin and orbital polarization
    for fixed $U$ and $J$. (b) spin and orbital polarization as a
    function of $J$ for $U-J=2.93$~eV and $T\approx{29}$~K. All data
    corresponds to RbO$_{2}$ input.}
  \label{fig:dmft-lowt-rbo2}
\end{figure}

Even though we do not attempt to fully resolve the resulting spin and
orbital patterns, we can make a number of interesting
observations. First of all, there are drastic differences between
RbO$_2$ (Fig.~\ref{fig:dmft-occiter}(a)) and the simple SC DOS
(Fig.~\ref{fig:dmft-occiter}(b)). The latter oscillates between three
states with different spin and orbital polarization (SP and OP) and is
insulating for $J\leq{0.8}$~eV, while for higher $J$ it is metallic
with no SP and OP (in these calculations both $U$ and $J$ have been
varied while keeping $U_\text{eff}=U-J$ constant). For RbO$_2$ we can
distinguish three different regimes (see
Fig.~\ref{fig:dmft-lowt-rbo2}(b)). For $J\leq{0.5}$~eV, the occupation
oscillates between three different states with different SP and almost
no OP. For $J=0.62$ and $0.8$~eV, a stable solution with large OP and
zero SP appears, while a further increase of $J$ induces a stable SP
and reduced OP. The system is insulating for all $J\leq{1.0}$~eV,
while for $J=1.2$~eV it is a ferromagnetic half-metal with full SP and
no OP. In Fig.~\ref{fig:dmft-lowt-rbo2}(a) we show the SP and OP of
RbO$_2$ as a function of temperature for the realistic values
$U=3.55$~eV and $J=0.62$~eV. The system is insulating and while
essentially no SP develops down to $T\approx{30}$~K, OP appears below
$T\approx{60}$~K and reaches almost its maximum at $T\approx{30}$~K.
While it is not possible from our calculations to make a prediction
about the character of the expected spin- and orbitally-ordered ground
state, the above temperatures are consistent with our previous
estimate of the ordering temperature based on total energy differences
of different orbitally ordered configurations obtained from DFT+$U$
calculations at $T=0$~K \cite{2009_kovacik}.

\section{Summary and Conclusions}

In summary our calculations clearly show that for realistic values of
the interaction parameters $U$ and $J$, RbO$_2$ at room temperature is
a paramagnetic Mott insulator without exhibiting any symmetry-breaking
long-range order. We find pronounced \emph{quantitative} differences
between the widely used SC DOS and the realistic electronic structure
of RbO$_2$, which leads to a strong asymmetry between the 1/4-filled
and 3/4-filled cases. We also find indications of complex spin and
orbital order below $T \approx 30$~K, the character of which seems to
depend strongly on $J$. Furthermore, at low temperature RbO$_2$
exhibits clear \emph{qualitative} differences compared to the
simplified SC DOS. It will be interesting to clarify in future work
whether single site DMFT is capable to resolve the complicated spin
and orbital patterns predicted within model calculations based on a
perturbative treatment of electron-electron interaction and a
simplified electronic structure of RbO$_2$
\cite{2010_ylvisaker,2011_wohlfeld}.

\acknowledgements

The authors are indebted to Emanuel Gull for his help with the
installation and general use of the CT-HYB code. This work was done
mostly within the School of Physics at Trinity College Dublin,
supported by Science Foundation Ireland under Ref.~SFI-07/YI2/I1051,
and made use of computational facilities provided by the Trinity
Center for High Performance Computing.

\bibliography{references}

\end{document}